\documentclass[aps,prb,twocolumn,superscriptaddress,showpacs,amsmath]{revtex4}
\usepackage{graphicx}

\begin{document}
\title{The enhancement of ferromagnetism in uniaxially stressed diluted
magnetic semiconductors}

\author{Yu.G. Semenov}
\email{ygsemeno@ncsu.edu}
\altaffiliation[Also at ]{Institute of
Semiconductor Physics, National Academy of Sciences of Ukraine,
Prosp. Nauki, 45, Kiev 03028 Ukraine}
\affiliation {Department of
Electrical and Computer Engineering, North Carolina State
University, Rayleigh, NC 27695-7911, USA}

\author{V. A. Stephanovich}
\homepage{http://cs.uni.opole.pl/~stef}
\email{stef@uni.opole.pl}
\affiliation {University of Opole, Institute of Mathematics
         and Informatics,Opole,45-052, Poland}
\date{\today }

\begin{abstract}
We predict a new mechanism of enhancement of ferromagnetic phase
transition temperature $T_c$ in uniaxially stressed diluted
magnetic semiconductors (DMS) of p-type. Our prediction is based
on comparative studies of both Heisenberg (inherent to undistorted
DMS with cubic lattice) and Ising (which can be applied to
strongly enough stressed DMS) models in a random field
approximation permitting to take into account the spatial
inhomogeneity of spin-spin interaction. Our calculations of phase
diagrams show that area of parameters for existence of
DMS-ferromagnetism in Ising model is much larger than that in
Heisenberg model.
\end{abstract}

\maketitle

\section{Introduction}

Recently (see \cite{Tanaka,Dietl} and references therein), great advances
have been made in the problem of ferromagnetism (FMm) of p-doped diluted
magnetic semiconductors (DMS). The question about influence of different
physical phenomena in DMS on the critical temperature $T_{c}$ of
ferromagnetic (FM) phase transition is of prime interest for searching for
the future study trends. After Pashitskii and Ryabchenko prediction of FMm
in DMS,\cite{PR} the competition between FM correlations mediated by
indirect long-range spin-spin interaction and direct short-range
antiferromagnetic (AFM) interaction is considered to be decisive in the
formation of FM state in DMS. In other words, to obtain FM state with high
enough $T_{c}$, we need to inhibit the AFM contribution.

Due to short-range character of AFM interaction, only close pairs of
magnetic ions contribute to it. This contribution can be minimized by
decreasing the magnetic ion concentration $n_i$. On the other hand, when $%
n_i $ is too small, ferromagnetism can be destroyed. To retain FM ordering
in this case, we may increase the carriers concentration $n_c$ (in p-type
DMS this corresponds to holes concentration $n_h$ ). The calculations of $%
T_c $ performed in a mean field approximation (MFA) \cite{PR} support this
standpoint predicting an increase of $T_c$ as $n_{c}^{1/3}n_{i}$.

However, at large $n_c$, the Friedel oscillations of carrier spin
polarization become significant so that MFA becomes inapplicable. More
thorough calculations with respect to Friedel oscillations (i.e. beyond MFA)
corroborate above statement and show that in DMS undergoing FM phase
transition $n_c$ cannot exceed some critical value related to $n_i$. The
reason for that is the oscillations of the RKKY interaction at the scale of $%
1/k_{F}\sim n_{c}^{-1/3}$, which makes impossible long-range FM correlations
if ${\overline{r}}k_{F}\geq 1$, (${\overline{r}}$ is an inter-ion mean
distance) \cite{SS}.

To properly account for above Friedel oscillations, which is indeed a
spatial dispersion of inter-ion interaction, we developed so-called random
field approximation (RFA) in Ref.\onlinecite{SS}. In that work, the ion-ion
interaction has been considered in the context of Ising model. This model
can be applied for axially symmetric semiconductors with magnetic ions
interacting indirectly via holes (i.e. with RKKY interaction). The reason
for Ising model usage for RKKY interaction is a complex structure of valence
band in DMS that picks out the hole angular momenta projections $J_Z=\pm 3/2$
for lowest heavy hole subbands in crystals with distorted cubic or uniaxial
lattice.

Additionally to spatial dispersion of inter-ion interaction, there are
fluctuations of local magnetic field direction that also cannot be described
in terms of MFA. This effect stems from the contribution of transversal spin
components in effective Heisenberg - like Hamiltonian of spin-spin
interaction. Namely, RKKY-interaction in p-doped undistorted DMS with cubic
lattice represents this situation.

In the present paper we pay attention to the fact, that transition from
Heisenberg spin-spin interaction to Ising one, i.e. exclusion of transversal
spin components from Heisenberg Hamiltonian decreases the system entropy and
therefore can enhance $T_c$. For quantitative description of this effect we
present a comparative analysis of the RFA-theories for critical temperature $%
T_c$ in Heisenberg and Ising models. This analysis has been made to
determine the role of directional fluctuations (inherent to Heisenberg
model) of localized spins in a random magnetic field. The uniaxial stresses
in typical DMS-structures grown on a substrate with some mismatch of lattice
constants is shown to be the factor responsible for appearance of Ising-like
interaction between magnetic ions spins. Thus, we predict a new effect
implying that increasing of $T_c$ can be controlled within certain bounds by
the uniaxial stresses in DMS. In other words, we expect that strain
engineering can efficiently control the value of ferromagnetic phase
transition temperature resulting from the hole-mediated exchange interaction
between magnetic ions in DMS.

\section{Theoretical background}

\subsection{Heisenberg model}

The Hamiltonian of Heisenberg model for DMS reads
\begin{equation}
{\cal H}=-\sum_{j<j'}J({\vec{r}}_{j,j'}){\vec{S}}_{j}{\vec{%
S}}_{j^{\prime }}+\sum_{j}{\vec{H}}_{0}{\vec{S}}_{j},  \label{mu0}
\end{equation}%
where external magnetic field ${\vec{H}}_{0}$ and interaction $J({\vec{r}}%
_{j,j'})$ is measured in energy units (i.e. $g\mu =1$, $\mu $ is
the Bohr magneton). The transition to Ising model means,
hereafter, keeping
only $S_{jZ}S_{j'Z}$ - components in scalar product ${\vec{S}}_{j}{%
\vec{S}}_{j'}$.

The Hamiltonian (\ref{mu0}) incorporates two kinds of randomness. First,
(the spatial disorder) is that spin can be randomly present or absent in the
specific $j$-th cite of a host semiconductor. Second, (the thermal disorder)
is a random quantum state of a spin in $j$-th cite. These spatial and
thermal fluctuations can be taken into consideration by introduction of
random field rather than mean field.

In the random field approximation, we consider every spin ${\vec{S}}_{j}$ as
a source of fluctuating (random) field
\begin{equation}
{\vec{H}}_{ri,j}=-J({\vec{r}}_{i}-{\vec{r}}_{j}){\vec{S}}_{j}  \label{mu1}
\end{equation}%
affecting other spin at the sites ${\vec{r}}_{i}$. In other words, every
spin is subjected to some random (rather then mean) field, created by all
other spins. So, all thermodynamic properties of the system will be
determined by the distribution function $f({\vec{H}}_{r})$ of the random
field ${\vec{H}}_{r}$. Namely, any spin dependent macroscopic quantity (like
magnetization) $<<A>>$ reads
\begin{equation}
<<A>>=\int <A>_{{\vec{H}}_{r}}f({\vec{H}}_{r})d{\vec{H}}_{r},  \label{mu2}
\end{equation}
where
\begin{equation}
<A>_{{\vec{H}}_{r}}=\frac{{\rm Tr}\left\{ A\exp \left( -{\cal H}%
_{Z}/T\right) \right\} }{{\rm Tr}\exp \left( -{\cal H}_{Z}/T\right) }
\label{mu3}
\end{equation}%
is a single particle thermal average with temperature $T$ and effective
Zeeman Hamiltonian ${\cal H}_Z={\vec H}_r{\vec S}$.

The distribution function $f({\vec H}_r)$ is defined as

\begin{equation}
f({\vec{H}}_{r})=\left\langle \overline{\delta \left( {\vec{H}}%
_{r}+\sum_{j(\neq i)}J({\vec{r}}_{i}-{\vec{r}}_{j}){\vec{S}}_{j}-{\vec{H}}%
_{0}\right) }\right\rangle ,  \label{fHf}
\end{equation}%
where the bar means averaging over spatial disorder. Our RFA approach is
based on micro-canonical statistical theory of magnetic resonance line shape.%
\cite{Stoneham} Latter theory assumes the additivity of local molecular
field contributions ${\vec{H}}_{r}=\sum_{j}$ ${\vec{H}}_{ri,j}$ of each
particle $j$ (Eq. (\ref{mu1})) as well as the non-correlative spatial
distributions of magnetic ions.

Latter assumptions with respect to spectral representation of $\delta $
function permit to transform Eq. (\ref{fHf}) to the non-linear integral
equation for $f({\vec{H}}_r)\equiv f({\vec{H}})$ in thermodynamic limit.
Introducing the probability $n_i({\vec{r})}d^3{\vec{r}}$ for small volume $%
d^3{\vec{r}}$ to be occupied by a particle, we obtain
\begin{subequations}
\label{eq:mas1}
\begin{eqnarray}
&&f({\vec{H}})=\int \exp \left[ i{\vec{\tau}}({\vec{H}}-{\vec{H}}_{0})-{\cal %
G}\left( {\vec{\tau}}\right) \right] \frac{d^{3}{\vec{\tau}}}{(2\pi )^{3}};
\label{eq5} \\
&&{\cal G}\left( {\vec{\tau}}\right) =\int_{V}\Psi ({\vec{r}})n_{i}({\vec{r}}%
)d^{3}{\vec{r},}  \label{eq5p} \\
&&\Psi ({\vec{r}}) \equiv <<1-\exp
[iJ({\vec{r}}){\vec{S}}{\vec{\tau}}]>>=\nonumber \\
&&=\int W({\vec{H})}f({\vec{H}})d^3H,  \label{eq5q} \\
&&W({\vec{H})} = <1-\exp [iJ({\vec{r}}){\vec{S}}{\vec{\tau}}]>_{{\vec{H}}}.
\label{eq5r}
\end{eqnarray}
\end{subequations}
Eqs (\ref{eq:mas1}) represent the integral equation for
distribution function $f({\vec{H}})$. In general case this
equation can be solved only numerically.

However, in many cases (e.g. for $T_{c}$ or magnetization calculations) it
is possible to avoid the solution of the integral equation since in these
cases it is exactly reducible to the set of transcendental equations for
macroscopic quantities like $<<{\vec{S}}^{n}>>$, $1\leq n\leq 2S$ of the
system. Simplest situation corresponds to the case $S=1/2$, where only
magnetization ${\vec{{\cal M}}}=-g\mu <<{\vec{S}}>>$ (or in dimensionless
units ${\vec{m}}=-2<<{\vec{S}}>>$; $g$ is g-factor of a magnetic ion) is a
unique order parameter to be found.

Thus, in the case of $H_0=0$ and ${\vec S}=\frac{1}{2}{\vec{\sigma}}$ ($%
\vec{\sigma}$ are the Pauli matrices) Eq. (\ref{eq5}) takes the form
\begin{eqnarray}
&&f({\vec{H}})=\int \exp \left[ i\vec{{\tau }}{\vec{H}}\right] \times \nonumber \\
&&\times \exp \left\{ -%
{\cal F}_{0}\left( \frac{\tau }{2}\right) -i\frac{\vec{\tau}}{\tau }\vec{m}%
{\cal F}_{1}\left( \frac{\tau }{2}\right) \right\} \frac{d^{3}{\vec{\tau}}}{%
(2\pi )^{3}},  \label{eq5h}
\end{eqnarray}
where
\begin{equation}
\vec{m}=\int \frac{\vec{H}}{H}\tanh \frac{H}{2T}f(\vec{H})d^{3}\vec{H};
\label{eq5a}
\end{equation}%
\begin{equation}
{\cal F}_{0}(x)=\int_{V}n({\vec{r}})\left[ 1-\cos \left( J(\vec{r})x\right) %
\right] d^{3}\vec{r};  \label{eq6a}
\end{equation}%
\begin{equation}
{\cal F}_{1}(x)=\int_{V}n(\vec{r})\sin \left( J({\vec{r}})x\right) d^{3}\vec{%
r}.  \label{eq7a}
\end{equation}%
Here $\tau =|{\vec{\tau}}|\equiv \sqrt{\tau _{x}^{2}+\tau _{y}^{2}+\tau
_{z}^{2}}$. To derive (\ref{eq5h}) we have used following property of Pauli
matrices:
\begin{equation}
\exp ({\vec{\sigma}}{\vec{b}})=\cosh b+\frac{{\vec{\sigma}}{\vec{b}}}{b}%
\sinh b,\quad b\equiv |{\vec{b}}|.  \label{eq7c}
\end{equation}

Substitution of equation (\ref{eq5h}) into Eq.(\ref{eq5a}) results in a
single closed equation for order parameter ${\vec{m}}$:
\begin{eqnarray}
&&{\vec m}=\frac 1{\left( 2\pi \right) ^3}\int d^3\vec{H}\int d^3{\vec{\tau}}%
\frac{\vec{H}}H\tanh \frac H{2T}\times \nonumber \\
&&\times \exp \left\{ -{\cal F}_0\left( \frac \tau 2%
\right) +i{\vec{\tau}}{\vec{B}}\right\} ,  \label{eq9a}
\end{eqnarray}
where
\begin{equation}
\vec{B}=\vec{H}-\vec{m}\frac{{\cal F}_1\left( \frac \tau 2\right) }\tau .
\label{eq10a}
\end{equation}

To simplify the vector equation (\ref{eq9a}), we scalarwise multiply its
left- and right-hand sides by ${\vec{m}}$ and integrate the resulting
equation for scalar quantity $m\equiv |\vec{m}|$ over the angle between
vectors $\vec{H}$ and $\vec{m}$. The final result (see Appendix A for
details of its derivation) reads
\begin{eqnarray}
&&m=-6\int_{0}^{\infty }{\cal B}_{1/2}^{H}(t)e^{-{\cal
F}_{0}(t/2T)}\times \nonumber \\
&&\times R_2%
\left( m{\cal F}_{1}\left( \frac{t}{2T}\right) \right) dt,  \label{eq24} \\
&&{\cal B}_{1/2}^{H}(t)=\frac{1}{3}{\rm csch}\pi t(1+\pi t\coth \pi t),\
\label{eq24b} \\
&&R_{n}(x)=\frac{x\cos x-\sin x}{x^{n}}.  \label{eq24r}
\end{eqnarray}%
Trivial solution $m=0$ of the equation (\ref{eq24}) corresponds to
paramagnetic phase. Under certain system parameters and
temperatures, the equation (\ref{eq24}) has nontrivial solution
that determines the phase transition to the state with spontaneous
magnetization.

To find the critical temperature $T_{c}$, we use the Landau theory with $m$
as an order parameter. For this purpose, we may derive (see Appendix B for
details of derivation) the free energy of the system in the form
\begin{eqnarray}
&&F^H=F_0+\frac{1}{2}m^{2}+6\int_{0}^{\infty }{\cal B}_{1/2}^{H}(t)e^{-%
{\cal F}_{0}(t/2T)}\times \nonumber \\
&&\times \frac{\sin \left( m{\cal F}_{1}\left( \frac{t}{2T}\right)
\right) }{m{\cal F}_{1}^{2}\left( \frac{t}{2T}\right) }dt.
\label{fenh}
\end{eqnarray}
In the vicinity of $T_{c}$, the free energy (\ref{fenh}) can be substituted
by a Landau expansion

\begin{eqnarray}
F_{L}^{H} &=&F_{0}+\frac{1}{2}m^{2}\left( 1-2A_{1}^{H}\right) +\frac{1}{20}%
m^{4}A_{3}^{H};  \label{q1} \\
A_{n}^{H} &=&\int_{0}^{\infty }{\cal B}_{1/2}^{H}(t)e_{1}^{-{\cal F}%
_{0}(t/2T)}{\cal F}_{1}^{n}\left( \frac{t}{2T}\right) dt,  \nonumber
\end{eqnarray}%
where $F_{0}$ is a system free energy in a paramagnetic phase.

It should be noted here that contrary to conventional phenomenological
Landau expansions of a free energy, the coefficients $A_n^H$ in the function
(\ref{q1}) have been derived microscopically within our RFA approach. Free
energy functions (\ref{fenh}) and (\ref{q1}) give a possibility to describe
the experimentally observable equilibrium thermodynamic characteristics
(like magnetic susceptibility, specific heat etc) of the DMS both in
paramagnetic and in ferromagnetic phases.

According to Landau theory of phase transitions, the phase transition
temperature $T_{c}$ is reached, when coefficient $1-2A_{1}^{H}=0$ in Eq. (%
\ref{q1}). This is because $T_{c}$ is defined as a temperature, where
nonzero infinitesimal magnetization appears. Obviously, the same equation
can be obtained from the Eq. (\ref{eq24}) for magnetization in the limit $%
m\to 0$. The explicit form of the equation for $T_c\equiv T_c^H$
reads

\begin{equation}
1=2\int_{0}^{\infty }{\cal B}_{S}^{H}(t){\cal F}_{1}\left( \frac{t}{%
2T_{c}^{H}}\right) e^{-{\cal F}_{0}\left( t/2T_{c}^{H}\right) }dt,
\label{eq27a}
\end{equation}%
where $S=1/2$ in our case.

Actually, the Eq. (\ref{eq27a}) determines the $T_{c}^{H}$ as an implicit
function of system parameters (like $n_i$, $n_c$ etc). This function can be
considered as a phase diagram that separates the region of parameters where
the ferromagnetic phase with $m \neq 0$ exists from that where $m=0$. Latter
phase may be paramagnetic or spin glass phase. In principle, our RFA method
permits to investigate this question. This study, however, is beyond the
scope of the present paper.

The limit $T_{c}^{H}\to 0$ in (\ref{eq27a}) gives the relation between
parameters of the system, which determines the condition for ferromagnetic
ordering to occur in DMS at $T=0$. The explicit form of this condition reads
\begin{equation}
1<\frac{4}{3\pi }\int_{0}^{\infty }\frac{{\cal F}_{1}\left( x\right) }{x}e^{-%
{\cal F}_{0}\left( x\right) }dx.  \label{cch}
\end{equation}%

\subsection{Ising model}

Let us consider now Ising model. In this case all effective magnetic fields
are directed along $OZ$ axis, so that scalar product reduces to ${\vec{S}}{%
\vec{\tau}}=S_{Z}\tau $, $S_{Z}=-1/2,1/2$ and Eq.(\ref{eq5}) becomes
\begin{eqnarray}
&&f(H)=\frac{1}{2\pi }\int_{-\infty }^{\infty }e^{iH\tau -{\cal G}(\tau
)}d\tau ,  \label{eq7} \\
&&{\cal G}(\tau )=\left\langle \left\langle 1-\int_{V}n_{i}(\vec{r})\left(
e^{-i\tau J(\vec{r})S_{Z}}\right) d^{3}{\vec{r}}\right\rangle \right\rangle =
\nonumber \\
&=&{\cal F}_{0}\left( \frac{\tau }{2}\right) +im{\cal F}_{1}\left( \frac{%
\tau }{2}\right) ,  \label{eqG}
\end{eqnarray}

where definition of $m$ is similar to Eq.(\ref{eq5a}):
\begin{equation}
m=\int_{-\infty }^\infty \tanh \left( \frac H{2T}\right) \ f(H)dH.
\label{eq7b}
\end{equation}
Multiplying Eq.(\ref{eq7}) by $\tanh (H/2T)$ and integrating over $H$, we
obtain the transcendental equation for order parameter $m$. The explicit
form of this equation reads
\begin{subequations}
\label{eq:ar2}
\begin{eqnarray}
&&m=\int_{-\infty }^\infty {\cal B}_{1/2}^I(t)\exp \left( -{\cal
F}_0\left( \frac t{2T}\right) \right) \times \nonumber \\
&&\times \sin \left(m{\cal F}_1\left( \frac t{2T}\right) \right)
dt, \label{eq8}
\end{eqnarray}

where
\begin{equation}
{\cal B}_{1/2}^I(x)=\frac 1{2\pi }\int_{-\infty }^\infty \tanh \left( \frac h%
2\right) \sin \left( xh\right) dh=\frac 1{\sinh \pi x}.  \label{eq8b}
\end{equation}
\end{subequations}
The equations for free energy and critical temperature $T_c$ can
be obtained similarly to those for Heisenberg model. They read
\begin{eqnarray*}
F^I &=&F_0+\frac 12m^2-\int_{-\infty }^\infty {\cal B}_{1/2}^I(\pi
t)\exp \left( -{\cal F}_0\left( \frac t{2T}\right) \right)\times \\
&\times &\frac{1-\cos \left( m%
{\cal F}_1\left( \frac t{2T}\right) \right) }{{\cal F}_1\left( \frac t{2T}%
\right) }dt, \\
F_L^I &=&F_0+\frac 12m^2\left( 1-A_1^I\right) +\frac 1{24}m^4A_3^I, \\
A_n^I &=&2\int_0^\infty {\cal B}_{1/2}^I(t)\exp \left( -{\cal F}_0\left(
\frac t{2T}\right) \right) {\cal F}_1^n\left( \frac t{2T}\right) .
\end{eqnarray*}
Similar to Eq. (\ref{eq27a}), the explicit form of equation for $T_c\equiv
T_c^I$ reads
\begin{equation}
1=2\int_0^\infty {\cal B}_{1/2}^I(t){\cal F}_1\left( \frac t{2T_c^I}\right)
\exp \left( -{\cal F}_0\left( \frac t{2T_c^I}\right) \right) dt,
\label{eq20}
\end{equation}
while the equation for FMm region in the phase diagram at $T=0$ has
following form
\begin{equation}
1<\frac 2\pi \int_0^\infty {\cal F}_1\left( x\right) e^{-{\cal F}_0\left(
x\right) }\frac{dx}x.  \label{cci}
\end{equation}

\section{Discussion of RKKY interaction}

\subsection{Non Gaussian fluctuations for $S=1/2$}

Let us analyze the equations for critical temperatures for Heisenberg (\ref%
{eq27a}) and Ising (\ref{eq20}) models in more details. The difference
between them consists only in the form of kernels of integrals for
Heisenberg (Eq. (\ref{eq24b})) and Ising (Eq. (\ref{eq8b})) cases, so that
equation for critical temperature in Heisenberg model can be transformed to
that in Ising model (and vice versa) by replacement of ${\cal B}_{1/2}^H(t)$
with ${\cal B}_{1/2}^I(t)$.

We start the analysis of these equations from their MFA asymptotics. To get
this asymptotics, the functions ${\cal F}_{0}(\xi )$ and ${\cal F}_{1}\left(
\xi \right) $ in the Eqs (\ref{eq6a}), (\ref{eq7a}) should be expanded up to
linear terms: ${\cal F}_{0}(\xi )\rightarrow 0$, ${\cal F}_{1}\left( \xi
\right) \rightarrow \xi {\overline{J}}$, where ${\overline{J}}=\int_{V}n_{i}(%
{\vec{r}})J({\vec{r}})d^{3}r$. After some algebra, the latter approximation
allows to reduce the Eqs(\ref{eq27a}), (\ref{eq20}) to the expressions for
critical temperatures in Heisenberg $T_{M}^{MF}(M=H)$ or Ising $(M=I)$
models:
\begin{equation}
T_{M}^{MF}=2{\overline{J}}\int_{0}^{\infty }{\cal B}_{1/2}^{M}(t)tdt=\frac{1%
}{4}\overline{J},  \label{eq28}
\end{equation}%
One can see that latter expression is identically the same to well-known MFA
result for $S=1/2$
\begin{equation}
T_{c}^{MF}=\frac{1}{3}S(S+1)\int_{V}n_{i}({\vec{r}})J({\vec{r}})d^{3}r.
\label{TcMF}
\end{equation}%
The Eqs (\ref{eq28}), (\ref{TcMF}) demonstrate also that MFA is independent
of the choice of Heisenberg or Ising model, $T_{H}^{MF}=T_{I}^{MF}=T_{c}^{MF}
$.

Next terms of expansion of the Eqs (\ref{eq6a}), (\ref{eq7a}) correspond to
Gaussian asymptotics for distribution function of local fields. The purpose
of subsequent analysis is to compare the (actual, i.e. non Gaussian)
fluctuations of longitudinal components of random field with those of
transversal ones.

Since our theory permits to find the distribution function $f(H)$
when a spatial dependence of $J({\vec{r}})$ is assigned, we should
specify a magnetic interaction in the system. Usually in the
problems of carrier-induced ferromagnetism in DMS, the RKKY
interaction\cite{KittelBook} is considered as an effective
spin-spin exchange interaction resulting in FM ordering. To
clarify the role of transversal spin fluctuations, here we use the
simplest possible form of the interaction and neglect all possible
factors that can influence $J(\vec{r})$ (such as nonparabolicity
of carrier dispersion law etc, see
Refs\onlinecite{Dietl01,McDon01,Mattis,Mauger,UmeKas} for more
details). Also, the stresses may change the form of $J(\vec{r}),$
see below for discussion.

In the case of simple one band carrier structure, the RKKY interaction reads
\begin{equation}
J(\vec{r})=-J_{0}x_{e}^{4/3}R_{4}(2k_{F}r),  \label{rkk}
\end{equation}%
where $x_{e}=n_{c}/N_{0}$, $J_{0}$ $=\left( \frac{3}{\pi }\right) ^{1/3}%
\frac{3}{2\hbar ^{2}}J_{ci}^{2}\Omega ^{2/3}m_{d}$, $J_{ci}$ is a
carrier-ion exchange constant, $N_{0}=1/\Omega $ is a concentration of the
cation cites, $m_{d}$ is the density of states effective mass. Note that in
our single band approximation, the effects of stress may influence on $x_{e}$
and $k_{F},$ see \cite{Dietl} for details. The threshold temperature of
ferromagnetic ordering in MFA now can be found by evaluation of integrals (%
\ref{TcMF}) with respect to (\ref{rkk}):
\begin{equation}
T_{c}^{MF}=\frac{1}{24\pi }J_{0}x_{i}^{4/3}\nu ^{1/3}.  \label{eq10}
\end{equation}%
Here, the factor $J_{0}x_{i}^{4/3}$ is independent of carrier concentration,
$x_{i}=n_{i}\Omega $ is a molar fraction of the magnetic ions. The ratio of
electron and magnetic ion concentrations $\nu =n_{c}/n_{i}=x_{e}/x_{i}$
plays a crucial role in our theory as a parameter separating the cases of
relatively small fluctuations with $\nu \ll \nu _{c}$ and that of large ones
with $\nu \approx \nu _{c}$; parameter $\nu _{c}$ is indeed a dimensionless
critical concentration (corresponding to equality sign in expressions (\ref%
{cch}) and (\ref{cci}) for Heisenberg and Ising models respectively).

The functions ${\cal F}_{0}(x)$ and ${\cal F}_{1}(x)$ in Eqs.(\ref{eq6a}), (%
\ref{eq7a}) (with respect to Eq.(\ref{rkk})) assume following form in the
case of homogeneous magnetic ions distribution, $n_{i}({\vec{r}})=n_{i}=const
$,
\begin{eqnarray}
{\cal F}_{0,1}\left( \xi \right)  &=&\varphi _{0,1}\left( \xi \right) /6\pi
\nu ,  \nonumber \\
\varphi _{0}\left( \xi \right)  &=&\int_{0}^{\infty }\left\{ 1-\cos \left(
\xi R_{4}(y)\right) \right\} y^{2}dy,  \label{s20a1} \\
\varphi _{1}\left( \xi \right)  &=&\int_{0}^{\infty }\sin \left( \xi
R_{4}(y)\right) y^{2}dy.  \label{s20b}
\end{eqnarray}

In the case of spin $S=1/2$, the result (\ref{eq10}) of MFA can be recovered
from Eqs (\ref{s20a1}), (\ref{s20b}) by their expansion up to linear terms $%
\varphi _{0}\left( \xi \right) \rightarrow 0$; $\varphi _{1}\left( \xi
\right) \rightarrow \xi $ (of course, with their further substitution into
Eq. (\ref{eq24}) or Eq. (\ref{eq8})). Gaussian asymptotics of distribution
function corresponds to the next term of expansion of the Eq. (\ref{s20a1}),
$\varphi _{0}\left( \xi \right) \rightarrow \pi \xi ^{2}/30,\ \varphi
_{1}\left( \xi \right) \rightarrow \xi $.

To account for real (non-Gaussian) distribution of fluctuating local field,
we do not expand Eqs (\ref{s20a1}),(\ref{s20b}) and calculate them
numerically.

In dimensionless variables, the equations for critical temperatures for both
above models assume following form
\begin{eqnarray}
&&\frac{1}{6\pi \nu }\int_{-\infty }^{\infty }{\cal B}_{1/2}^{M}(t)%
\varphi _{1}\left( \frac{t}{2\theta _{M}}\right) {\cal E}_{1/2}\left( \frac{t%
}{\theta _{M}}\right) dt=1, \nonumber  \\
&&{\cal E}_{1/2}\left( \frac{t}{\theta _{M}}\right) =\exp \left( -\frac{1%
}{6\pi \nu }\varphi _{0}\left( \frac{t}{2\theta _{M}}\right)
\right) , \label{s19}
\end{eqnarray}%
where $\theta _{M}=T_{c}^{M}/(J_{0}x_{e}^{4/3})$; $M$ stands for $H$
(Heisenberg model) or $I$ (Ising model). The result of calculation of $%
T_{c}^{M}/T_{c}^{MF}$ with the help of Eq. (\ref{s19}) as a
function of $\nu $ is reported in the Fig.1a. It is seen, that
there are curves that separate the areas of system parameters
(including temperature) where FM or non-FM phases occur.

\begin{figure}
\vspace*{15mm}
\centering{\
\includegraphics[width=9cm]{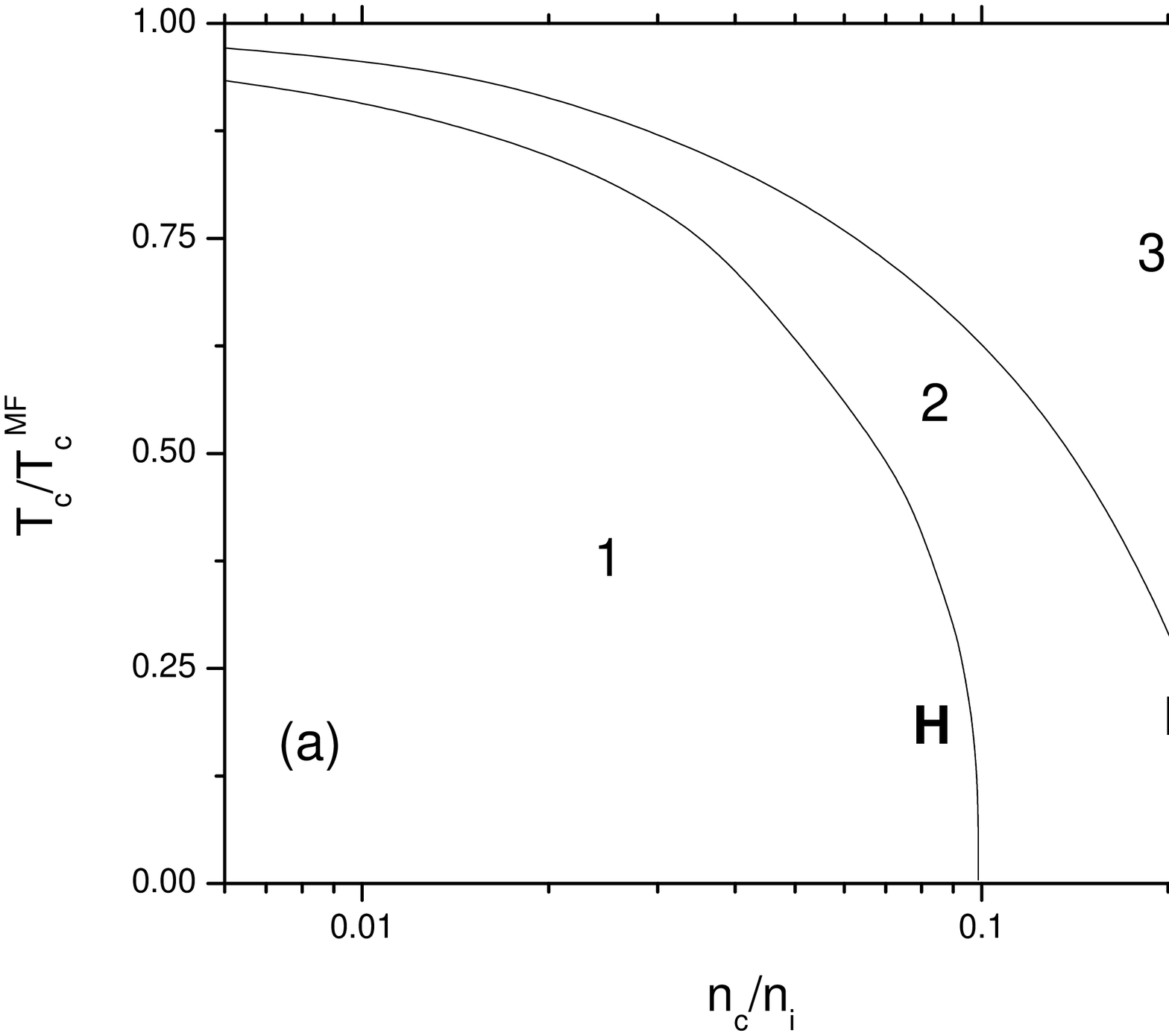}}
\centering{\
\includegraphics[width=9cm]{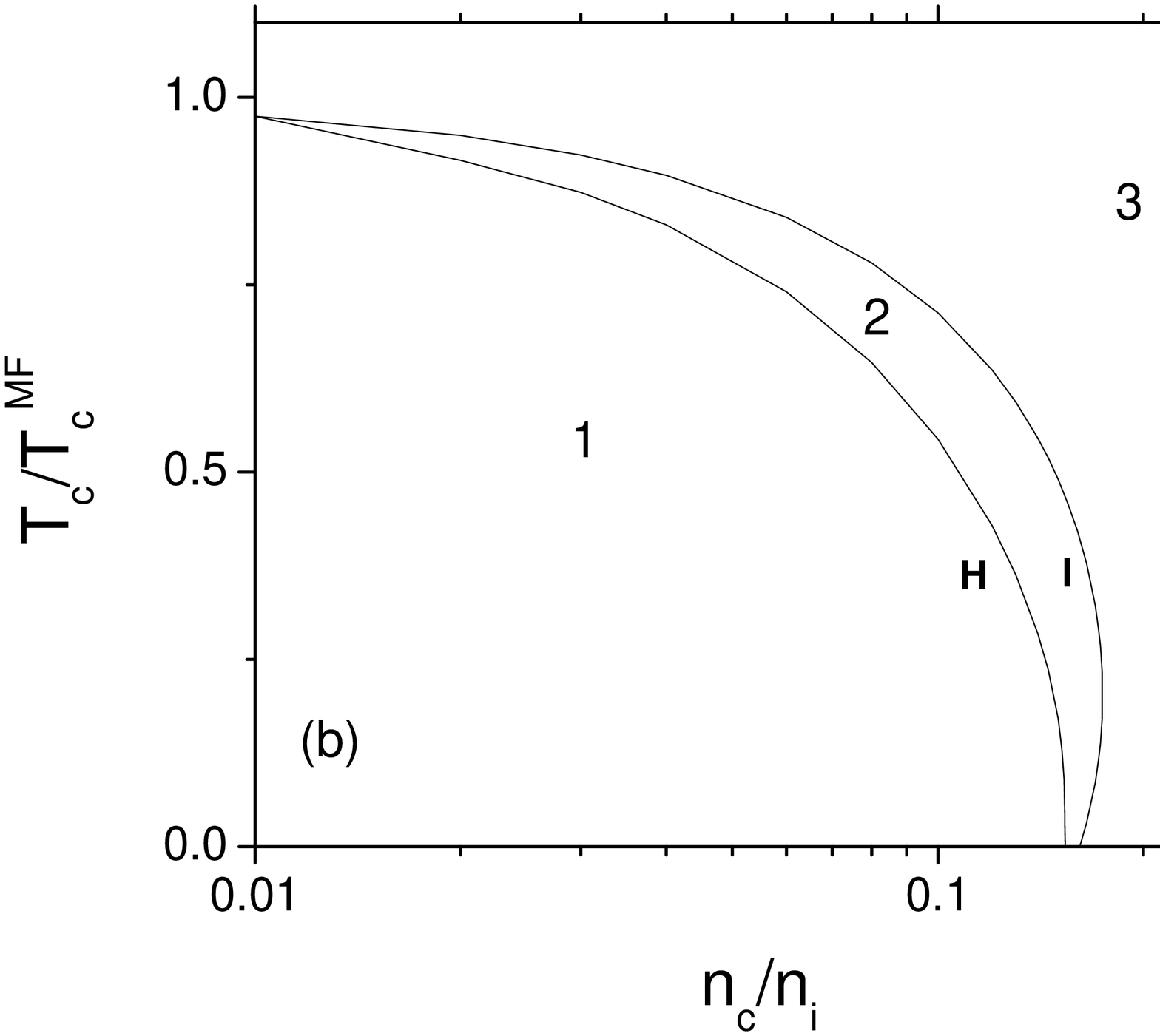}}
\vspace*{-2cm}
\caption{The phase diagram of the systems under
consideration. a) general (non-Gaussian) case for spin 1/2, b)
Gaussian approximation for spin 5/2. H - Heisenberg model, I -
Ising model. Region 1 corresponds to FM state for both models,
region 2 corresponds to FM state for Ising model and non FM state
for Heisenberg model, region 3 corresponds to non FM state for
both models. }
\end{figure}

Our results show the limited area of concentrations $0<\nu <\nu
_c$ which allow FM ordering in both considered models. So, we have
found $\nu _c=0.2473 $ for the Ising model and $0.0989$ for the
Heisenberg model. Our results also show the critical character of
dependence $T_c=T_c(\nu )$ that can be well approximated by the
function
\begin{equation}
T_c\simeq T_c^{MF}\left( 1-\nu /\nu _c\right) ^\lambda =\frac{J_0}{24\pi }%
x_i^{4/3}\nu ^{1/3}\left( 1-\nu /\nu _c\right) ^\lambda  \label{ss12}
\end{equation}
with $\lambda \simeq 0.47$ and $\lambda \simeq 0.63$ for Heisenberg and
Ising models respectively.

Thus, the fluctuations of transversal spin components suppress a
tendency towards FM ordering in the range of concentration ratios
$0<\nu <0.099$. Moreover, in the range $0.099<\nu <0.247$, our
results predict impossibility of FMm in DMS with Heisenberg-like
spin-spin interaction while in this interval of $\nu $ FMm can
still occur in DMS with Ising- like interaction between spins.
Since Ising model is inherent to uniaxially stressed
semiconductors, latter conclusion means that uniaxial distortion
can effectively inhibit transversal spin fluctuations thus
enhancing $T_c$.

\subsection{Gaussian fluctuations for $S=5/2$}

The equation (\ref{eq27a}) for critical temperature $T_c$ is exact in the
framework of our model, which means that it incorporates non-Gaussian
fluctuations. At the same time it is restricted by the case of ion spins $%
S=1/2$. Practically important case of $Mn^{2+}$ ions with $S=5/2$ needs a
special consideration. Mathematically, the problems arise for $S>1/2$, when
exponential function of spin operator is no more a linear function so that
Eq. (\ref{eq7c}) is no more valid. In this case we should use Sylvester
theorem for exponent of a Hermitian matrix $(2S+1)\times (2S+1)$(see, e.g.%
\cite{Korn} for more details). Appendix C presents the final result of such
calculations for $W({\vec H})$ and $S=5/2$ (Eq.(\ref{eq5r})).

It is apparent from Appendix C that the case $S=5/2$ involves much more
order parameters (such as $a_{m,n}=\left\langle c^{m}\eta _n\right\rangle
=\int c^{m}\eta _{n}f({\vec H})d^3H$, $m=0,2,4$; $n=0,1,2$, and $%
b_{m,n}=\left\langle c^{m}\phi _n\right\rangle $, $m=1,3,5$;
$n=0,1,2$) than in the case $S=1/2$. Not all of them play a
crucial role in a formation of FM phase in DMS. For example, if
the random field fluctuations are almost Gaussian, we can find the
expectation values of spin operators (\ref{mu2}) with the aid of
only two parameters, representing first and second moments of the
distribution function $f(H)$.

Our calculations show that Gaussian approximation for $f(H)$
adequately describes the actual phase diagram of DMS (except for
close vicinity of critical concentration which is not practically
important). That is why the analysis of the case of $S=5/2$ can be
performed with sufficient accuracy in Gaussian approximation. To
do this, the Eq. (\ref{ac6}) should be substituted by its
expansion in $t=J({\vec r}){\tau }$ up to the second order. The
result reads
\begin{widetext}
\begin{equation}
W({\vec H}) =ic\left( 3\coth \frac{3H}{T}-\frac{1}{2}\coth \frac{H}{2T}%
\right) J({\vec{r}}){\tau } +\left( \frac{5}{4}\left(
1+4c^{2}\right) -\frac{\left( 3c^2-1\right)
\left( 1+4\cosh \frac{H}{T}\right) }{1+2\cosh \frac{2H}{T}}\right) \frac{%
\left( J({\vec{r}}){\tau }\right) ^{2}}{2}.\label{HG1}
\end{equation}
\end{widetext}
We apply this result to obtain the equation for critical temperature. The
magnetization and mean value of random field ${\vec H}$ are negligibly small
at this temperature that suggests that $f({\vec H})$ is isotropic. Thus,
performing angular averaging over the directions of ${\vec H}$ in Eq.(\ref%
{eq5q}), we can put $c^2=1/3$ in the Eq. (\ref{HG1}).

To relate this result to the case $S=1/2$, we note that first term of Eq. (%
\ref{HG1}) includes Brillouin function for spin $S=5/2$
\begin{equation}
B_{S}\left( \frac{SH}{T}\right) =(1+\frac{1}{2S})\coth \frac{(S+1/2)H}{T}-%
\frac{1}{2S}\coth \frac{H}{2T}.  \label{HGa}
\end{equation}
After simple algebra, the Eq.(\ref{eq5q}) takes following form for
arbitrary spin $S$
\begin{equation}
\Psi ({\vec r})=i\frac{\vec \tau }{\tau }{\vec m}SJ({\vec r})+\frac 13 S(S+1)%
\frac{\left( J({\vec r}){\tau }\right) ^{2}}{2},  \label{HG2}
\end{equation}

where
\begin{equation}
\vec{m}=\int \frac{\vec{H}}{H}B_{S}\left( \frac{SH}{T}\right) f(\vec{H})d^3H.
\label{HG3}
\end{equation}

One can see that the above equations are formally similar to those for $%
S=1/2 $ if we expand Eqs (\ref{eq6a}) and (\ref{eq7a}) up to first
nonvanishing terms. The Eq. (\ref{HG2}) determines the components of Fourier
image of distribution function (\ref{eq5h}) in Gaussian approximation
\begin{eqnarray}
&&{\cal F}_0\left(\frac{\tau }{2}\right)=\frac{1}{3}S(S+1)\overline{J^{2}}%
\frac{\tau ^{2}}{2};\quad {\cal F}_{1}\left(\frac{\tau }{2}\right)=S%
\overline{J}\tau ,  \nonumber \\
&&\overline{J^n}=\int_{V}n(\vec{r})J^{n}({\vec{r}})d^3r.
\label{HG4}
\end{eqnarray}

Equations (\ref{HG3}) and (\ref{HG4}) reduce the problem of $T_c$
determination in Gaussian approximation for arbitrary ion spin to the case
of non-Gaussian fluctuations for spin $S=1/2$ considered above. Namely,
after substitution of these equations to Eq. (\ref{eq27a}) along with proper
generalization (for the case of arbitrary spin) of the function ${\cal B}%
_{1/2}^{H}(t)$, we can use this equation to find $T_c$ for any $S>1/2$.
Calculations performed in the same manner as Eqs. (\ref{ap4}) and (\ref{ap4a}%
) yield
\begin{widetext}
\begin{equation}
{\cal B}_{S}^{H}(t)=\frac{1}{6S}\left( \coth \frac{\pi t}{2S+1}-\coth \pi
t\right) +\frac{\pi t}{6S}\left( \frac{{\rm csch}^2\frac{\pi t}{2S+1}}{2S+1}-%
{\rm csch}^2\pi t\right) .  \label{HG5}
\end{equation}
\end{widetext}
It can be readily shown that in the case $S=1/2$ the
Eq. (\ref{HG5}) reduces to simpler form (\ref{eq24b}).

Substitution of (\ref{HG4}) to (\ref{cch}) permits to obtain the necessary
condition for FMm formation at zeroth temperature in the form (see also Eqs.
(\ref{cch}) and (\ref{cci}))
\begin{equation}
\overline{J}>\sqrt{\frac{3\pi }{8}\frac{S+1}{S}\overline{J^{2}}}.
\label{HG6}
\end{equation}%
This inequality has clear physical meaning: small dispersion and positive
(FM) inter-ion spin-spin interaction favors creation of FMm in DMS. Also,
large spin is also preferable for FMm formation due to relatively small
(quantum) fluctuations of its transversal components.

For comparison, we also consider Ising model for $S\geq 1/2$ in Gaussian
approximation (despite the possibility to account for non-Gaussian
fluctuations of random field for arbitrary $S\geq 1/2$).\cite{SS} Expansion
of the function ${\cal G}(\tau )$ (Eq.(\ref{eqG})) with arbitrary $S$ and $%
m=\int_{-\infty }^{\infty }{\rm B}_{S}\left( \frac{SH}{T}\right) \ f(H)dH$
up to second order leads to the distribution function in the form (\ref{eq7}%
) with ${\cal G}(\tau )$ in the form (\ref{eqG}), where
\begin{equation}
{\cal F}_{0}\left( \frac{\tau }{2}\right) =M_{2}\overline{J^{2}}\frac{\tau
^{2}}{2};\quad {\cal F}_{1}\left( \frac{\tau }{2}\right) =S{\overline{J}}%
\tau .  \label{IG1}
\end{equation}%
Parameters
\begin{equation}
m=\int_{-\infty }^{\infty }{\rm B}_{S}\left( \frac{SH}{T}\right) \ f(H)dH
\label{IGb}
\end{equation}%
and
\begin{eqnarray}
&&M_2=\left\langle \left\langle S_{Z}^{2}\right\rangle
\right\rangle =\int_{-\infty }^{\infty }{\rm Q}_{S}\left(
\frac{H}{T}\right) \ f(H)dH,
\label{IG2} \\
&&{\rm Q}_S\left( \frac{H}{T}\right) =S(S+1)-\nonumber \\
&&-S\coth \left( \frac{H}{2T}%
\right) {\rm B}_{S}\left( \frac{SH}{T}\right) ,  \label{IG3}
\end{eqnarray}%
should be found self-consistently with respect to Eqs. (\ref{eq7}), ( \ref%
{eqG}) and (\ref{IG1}). Thus, in Gaussian approximation, Ising model needs
self-consistent determination of two parameters, $\left\langle \left\langle
S_{Z}\right\rangle \right\rangle $ and $\left\langle \left\langle
S_{Z}^{2}\right\rangle \right\rangle $, while in the case of Heisenberg
model single parameter $<<|{\vec{S}}|>>$ needs to be determined. This is
because for Heisenberg-like interaction $<<{\vec{S}}^{2}>>=S(S+1)=const$
(Eq. (\ref{HG4})) so that this parameter depend neither on temperature nor
on field distribution.

The critical temperature can be found from the Eqs (\ref{IGb}) and (\ref{IG2}%
) as $m\to 0$
\begin{eqnarray}
&&1=\frac{\overline{J}\sqrt{8}}{\sqrt{\pi
M_2\overline{J^{2}}}}\times \nonumber \\
&&\times \int
_0^{\infty }xS{\rm B}_{S}\left( \frac{3\sqrt{2M_{2}\overline{J^{2}}}} {S(S+1)%
\overline{J}}\frac{x}{ \theta }\right) e^{-x^2}dx;  \label{IG5} \\
&&M_2=\frac{2}{\sqrt{\pi }}\int _0^{\infty }{\rm Q}_{S}\left( \frac{3\sqrt{%
2M_2 \overline{J^{2}}}}{S(S+1)\overline{J}}\frac{x}{\theta }\right)
e^{-x^{2}}dx,  \label{IG6}
\end{eqnarray}%
where $\theta =T_c/T_c^{MF}$ is a ratio of actual critical temperature and
that obtained in MFA approximation (Eq. (\ref{TcMF})). In this
approximation, the necessary condition to form FM state at $T=0$ in DMS has
the form of following inequality
\begin{equation}
\overline{J}>\sqrt{\frac{\pi }{2}\overline{J^{2}}}.  \label{IG7}
\end{equation}
The independence of Eq. (\ref{IG7}) of spin can be thought of (see Eq. (\ref%
{HG6}) for comparison) as a lack of transversal spin components contribution
to random field in Ising model.

Now we are ready to compare Heisenberg and Ising models for specific case of
RKKY interaction (\ref{rkk}). In Gaussian approximation we have to evaluate
only two integrals
\begin{equation}
{\overline{J}}=\frac{1}{6\pi }J_{0}x_{i}^{4/3}\nu ^{1/3},\,\overline{J^{2}}=%
\frac{2\pi ^{2}}{5}\nu \left( {\overline{J}}\right) ^{2}.  \label{IG8}
\end{equation}%
Note that the equations (\ref{IG8}) demonstrate the relation between Friedel
oscillations and first ($\propto {\overline{J}}$) and second ($\propto {%
\overline{J^{2}}}$) moments of distribution function of random magnetic
fields. Namely, at frequent Friedel oscillations (i.e. large $n_{c}$ and $%
\nu $) the influence of dispersion $({\overline{J^{2}}})^{1/2}$ (which is
"responsible" for disorder in the system) prevail over trend to order the
system due to mean value ${\overline{J}}$.\cite{SS} This means that frequent
Friedel oscillations at the scale of mean inter-ion distance inhibit
ferromagnetism in DMS.

Substitution of (\ref{IG8}) into (\ref{IG5}), (\ref{IG6}) permits
to find the dependence $\theta (\nu )$ for practically important
case $S=5/2$ (see Fig.1b). One can see that qualitatively
situations for $S=5/2$ and $S=1/2$ are similar: the region of FM
state is significantly larger for Ising model than that for
Heisenberg model. But there is also a difference. Comparison of
Fig.1a and 1b shows that the area between curves $T_c(n_c/n_i)$
for
Heisenberg (H) and Ising (I) models at low temperatures is smaller for $%
S=5/2 $. This means that for $S=5/2$ quantum fluctuations (which
is the only possible fluctuations at $T=0$) are inhibited as
compared to the case $S=1/2$.

This is actually a reflection of the well known fact that the
larger the magnitude of the quantum number of the spin, the "more
classical" it is, i.e. the smaller is the contribution of quantum
fluctuations of its transversal components. At $T \neq 0$
additional thermal fluctuations appear. It is seen from the
Figure, that at $T \leq T_c$ the
extension of FM phase due to enhancement of $T_c$ is almost the same for $%
S=5/2$ and $S=1/2$. This means that the thermal fluctuations of the spin do
not sensitive to its value.

\section{Conclusions}

In this paper we have proposed a new mechanism of the enhancement of FM
phase transition temperature $T_{c}$ by the uniaxial distortion of DMS. This
prediction is based on comparative analysis of Heisenberg (inherent to
undistorted DMS) and Ising (inherent to uniaxially distorted DMS) models.
The analysis of above models has been carried out in the framework of our
recently developed formalism, \cite{SS} random field method. This method,
which can be regarded as a substitution of conventional MFA for disordered
systems with given $J(\vec{r})$, permits to derive self-consistently the
equations for order parameter $m$ and the free energy functions of DMS.

Now we discuss in more details the influence of stresses on magnetic spin
Hamiltonian of DMS (\ref{mu0}). We consider such influence in two steps. In
first step we consider the influence of the stress on the effective
spin-spin interaction potential and in second step we consider the operator
part of the Hamiltonian.

Since the effective potential of indirect spin-spin interaction
strongly depends on the band structure of specific semiconductor
sample, which is affected by the stress, this potential by itself
may also depend on stress. This influence \cite{Dietl} is
manifested via both density of states (in our case we use the
effective mass of density of states $m_{d}$ (\ref{rkk})) at the
Fermi level and concentration of free holes, related to pinning of
the Fermi level by defects and impurities of different nature. As
it was noted in Ref. \cite{Dietl01}, the influence of pressure on
the density of states is small. The influence via pinning centers
depends entirely on their nature. We can imagine the situation
when the influence of pressure on the concentration is also small.
For example, in the cases of the absence (or small number) of
pinning centers or "synchronous movement" of pinning centers with
the valence band edge shift due to deformation potential, this
effect is negligible and our mechanism of influence of the
pressure will be decisive. Here we would like to emphasize that
there are no general problems to incorporate the possible
dependence of the concentration on the stress
into our consideration (this is simply one more modification of $J({\vec{r}}%
_{j,j'})$ in (\ref{mu0})). If we do so, the considered effect of
elimination of transversal spin components by the stress, which
has not been discussed in the literature, is an additional factor
enhancing $T_c.$

To discuss the stress effect on the operator part, we note that in the
typical case of p-type DMS with cubic lattice the transversal spin-spin
interaction can be substantially inhibited by the uniaxial stress of a
crystal. Really, such stress splits the valence band edge to light and heavy
hole subbands. The resulting heavy hole (HH) states are characterized by
angular momentum projection $J_Z=\pm 3/2$. For such states, the spin-flip
scattering of these holes on magnetic ions is forbidden due to angular
momentum conservation. Hence only longitudinal (along the distortion axis) $%
Z $-components of the spins will be present in the resulting indirect
spin-spin interaction via above heavy holes.

However, aforementioned spin-flip processes are not forbidden both for light
hole (LH) states (with spin projection $\pm 1/2$) and for transitions
between LH and HH states thus contributing to the indirect interaction of
transversal spin components of magnetic ions. Hence, the deformational
splitting of a valence band edge, leading to preferential occupancy of the
heavy hole subbands gives the anisotropy of indirect spin-spin interaction
in the form ${\cal H}=\sum_{i>j}\left[ J_{\parallel
i,j}S_{Zi}S_{Zj}+J_{\perp i,j}\left( S_{Xi}S_{Xj}+S_{Yi}S_{Yj}\right) \right]
$. The ratio $\gamma =J_{\perp i,j}/J_{\parallel i,j}$ ($0<\gamma <1$)
should be monotonically decreasing function of the ratio of heavy holes
concentration $n_{HH}$ to their complete concentration $n_{H}$. Thus, if
deformational HH - LH splitting exceeds the Fermi energy of the holes, the
ratio $n_{HH}/n_{H}$ reaches its maximal value ($n_{HH}/n_{H}\rightarrow 1$%
), corresponding to Ising Hamiltonian ($\gamma \rightarrow 0$) of the
indirect spin-spin interaction. In this case the region of ferromagnetic
state of DMS expands substantially towards both higher carriers (holes)
concentration and higher temperatures.

The detailed theoretical description of all above effects, which
is intimately related to the parameters of specific DMS sample and
experimental conditions, can be developed within the framework of
presented theory for any particular case. However, such
calculations can be done only numerically. Note, that the
magnitude of effect which we predicted is very sensitive to the
holes concentration $n_{h}$. The problem of its correct
determination, to the best of our knowledge, is related to the
influence of anomalous Hall effect. Latter effect lowers
reliability of $n_{h}$ determination from Hall effect data
\cite{Ohno1,Ohno2}.

For better illustration of our effect, we estimate now the magnitude of $%
T_{c}$ increase for typical ferromagnetic DMS $Ga_{1-x}Mn_{x}As$ deposited
on $GaAs$ and $GaP$ substrates. The mismatch $\Delta a$ of lattice constant $%
a$ leads to biaxial strain that splits the valence band with deformation
potential $b=-1.7$ eV by the value $\delta E_{1,2}=2\left\vert b\epsilon
_{zz}^{\prime }\right\vert $,\cite{BirPikus} where $zz$- component of strain
tensor $\epsilon _{zz}^{\prime }=-2(\Delta a/a)c_{12}/c_{11}$, and the ratio
of elastic moduli in GaAs $c_{12}/c_{11}=0.453$. For $x=0.035$ (or
concentration $n_{i}=7.76\cdot 10^{20}$ cm$^{-3}$) , the relative mismatch $%
\left\vert \Delta a/a\right\vert =0.002$ for $GaAs$ substrate and $%
\left\vert \Delta a/a\right\vert =0.036$ for $GaP$ substrate.\cite{Dietl01}
We can see that for $GaAs$ substrate small valence band splitting $\delta
E_{1,2}\simeq 6$ meV cannot suppress interaction of transversal $Mn$-spin
components for typical concentration $n_{c}=10^{20}$ cm$^{-3}$ which
corresponds to Fermi energy $\varepsilon _{F}\approx 80$ meV, whereas for $%
GaP$ substrate $\delta E_{1,2}\simeq 109$ meV $>\varepsilon _{F}$. Thus, our
mechanism predicts the enhancement of $T_{c}$ for $Ga_{1-x}Mn_{x}As$ on $GaP$
substrate by the factor 1.64 (see Fig.1b for $\nu =n_{c}/n_{i}=0.13$) as
compared to the same DMS but on $GaAs$ substrate.

Let us finally note that in the present paper we considered the enhancement
of $T_c$ due to RKKY interaction only. But there are also other mechanisms,
which can lead to appearance of ferromagnetism in DMS, see \cite{Dietl} for
details. These mechanisms will be eventually reduced to the Hamiltonian (\ref%
{mu0}) with modified potential $J({\vec{r}}_{j,j^{\prime}})$. Thus for
quantitative discussion of these mechanisms it is sufficient to substitute
the corresponding modified potential to our self-consistent equations.

\appendix
\section{}
We begin with equation (\ref{eq9a}) for magnetization.
\begin{widetext}
\begin{equation}
\vec{m}=\frac{1}{\left( 2\pi \right) ^{3}}\int d^{3}\vec{H}\int d^{3}{\vec{%
\tau}}\frac{\vec{H}}{H}\tanh \frac{H}{2T}\exp \left\{ -{\cal F}_{0}\left(
\frac{\tau }{2}\right) +i{\vec{\tau}}{\vec{B}}\right\} ,  \label{eq9aa}
\end{equation}%
First, we pass to spherical system for $\vec{{\bf \tau }}$ and integrate the
Eq.(\ref{eq9aa}) over $\vec{{\bf \tau }}$ directions. This yields
\begin{eqnarray}
\vec{m} &=&\frac{1}{2\pi ^{2}}\int d^{3}\vec{H}\int_{0}^{\infty }\tau
^{2}d\tau \frac{\vec{H}}{H}\tanh \frac{H}{2T}\exp \left[ -{\cal F}_{0}\left(
\frac{\tau }{2}\right) \right] \frac{\sin B\tau }{B\tau },  \label{eq13} \\
B &=&|{\vec{B}}|=\sqrt{H^{2}-2\cos \theta mH{\cal F}_{1}(\tau /2)/\tau
+\left( m{\cal F}_{1}(\tau /2)/\tau \right) ^{2}},  \label{e13a}
\end{eqnarray}%
where $\theta $ is an angle between vectors $\vec{m}$ and $\vec{H}$.

Next step is a scalar multiplication of its both sides by ${\vec{m}}$ that
yields
\[
m^{2}=\frac{1}{2\pi ^{2}}\int d^{3}\vec{H}\int_{0}^{\infty }\tau ^{2}d\tau
\frac{\vec{H}\vec{m}}{H}\tanh \frac{H}{2T}\exp \left[ -{\cal F}_{0}\left(
\frac{\tau }{2}\right) \right] \frac{\sin B\tau }{B\tau }
\]%
or
\begin{equation}
m=\frac{1}{2\pi ^{2}}\int d^{3}\vec{H}\int_{0}^{\infty }\tau ^{2}d\tau \cos
\theta \tanh \frac{H}{2T}\exp \left[ -{\cal F}_{0}\left( \frac{\tau }{2}%
\right) \right] \frac{\sin B\tau }{B\tau }  \label{ap1}
\end{equation}%
The rotational invariance of scalar product permits to point $\vec{H}$ along
$z$ axis and integrate over angular variables in (\ref{ap1}). This yields
\[
m=\frac{1}{2\pi ^{2}}\int_{0}^{2\pi }d\varphi \int_{0}^{\pi }\sin \theta
d\theta \int_{0}^{\infty }H^{2}dH\int_{0}^{\infty }\tau ^{2}d\tau \cos
\theta \tanh \frac{H}{2T}\exp \left[ -{\cal F}_{0}\left( \frac{\tau }{2}%
\right) \right] \frac{\sin B\tau }{B\tau },
\]%
where $B$ is defined by (\ref{e13a}). Consider
\[
I=\frac{1}{\tau }\int_{0}^{2\pi }d\varphi \int_{0}^{\pi }\sin \theta \cos
\theta \frac{\sin B\tau }{B}d\theta .
\]%
Change of variables
\[
\cos \theta =z=\frac{H^{2}+\left( m{\cal F}_{1}(\tau /2)/\tau \right)
^{2}-B^{2}}{2mH{\cal F}_{1}(\tau /2)/\tau }
\]%
with the help of (\ref{e13a}) reduces it to the form
\begin{eqnarray}
I &=&-\frac{\pi }{\tau \left( mH{\cal F}_{1}(\tau )/\tau \right) ^{2}}%
\int_{H+m{\cal F}_{1}(\tau )/\tau }^{H-m{\cal F}_{1}(\tau )/\tau }\left[
H^{2}+\left( m{\cal F}_{1}(\tau /2)/\tau \right) ^{2}-B^{2}\right] \sin
B\tau dB=  \nonumber \\
&=&4\pi R_{2}\left( m{\cal F}_{1}(\tau /2)\right) R_{2}(H\tau ),\ R_{n}(x)=%
\frac{x\cos x-\sin x}{x^{n}}.  \label{ap2}
\end{eqnarray}%
Substitution of (\ref{ap2}) into (\ref{ap1}) gives
\begin{equation}
m=\frac{2}{\pi }\int_{0}^{\infty }dH\int_{0}^{\infty }d\tau \tanh \frac{H}{2T%
}\exp \left[ -{\cal F}_{0}\left( \frac{\tau }{2}\right) \right] R_{2}\left(
mH_{1}(\tau /2)\right) R_{0}(H\tau ).  \label{ap3}
\end{equation}
\end{widetext}
It is also possible to integrate over $H$ in (\ref{ap3})
\begin{equation}
\int_{0}^{\infty }\tanh \frac{H}{2T}R_{0}(H\tau )\ dH=-3\pi T{\cal B}%
_{1/2}^{H}(t),\ t=\tau T.  \label{ap4}
\end{equation}%
\begin{equation}
{\cal B}_{1/2}^{H}(t)=\frac{\sinh \pi t+\pi t\cosh \pi t}{3\sinh ^{2}\pi t}.
\label{ap4a}
\end{equation}%
With respect to substitution $\tau T=t$ this gives final equation for $m$ in
the form (\ref{eq24}) of the text.

\section{}
We start the derivation of the free energy from the equation (\ref{eq24})
for magnetization (order parameter). We rewrite it in the form
\begin{equation}
m+6\int_0^\infty {\cal B}_{1/2}^H(\pi t)e^{-{\cal F}_0(t/2T)}R_2\left( m%
{\cal F}_1\left( \frac t{2T}\right) \right) dt=0.  \label{ap5}
\end{equation}
Now we recollect that if we have a free energy $F$ of a system, then the
equation for order parameter (in our case Eq. (\ref{ap5})) should minimize
it. In other words, Eq. (\ref{ap5}) should be equivalent to condition
\begin{equation}
\frac{\partial F}{\partial m}=0.  \label{ap6}
\end{equation}
Condition (\ref{ap6}) is a simple differential equation for $F$,
its solution yields
\begin{widetext}
\begin{eqnarray}
F &=&\int dm\left\{ m+6\int_0^\infty {\cal B}_{1/2}^H(\pi t)e^{-{\cal F}%
_0(t/2T)}R_2\left( m{\cal F}_1\left( \frac t{2T}\right) \right) dt\right\} =
\nonumber \\
&=&F_0+\frac 12m^2+6\int_0^\infty {\cal B}_{1/2}^H(\pi t)e^{-{\cal F}%
_0(t/2T)}\frac{\sin \left( m{\cal F}_1\left( \frac t{2T}\right) \right) }{m%
{\cal F}_1^2\left( \frac t{2T}\right) }dt.  \label{ap7}
\end{eqnarray}
This is indeed equation (\ref{fenh}) from the text.

To get Landau expansion of (\ref{ap7}), we simply expand $\sin \left( m{\cal %
F}_{1}\left( \frac{t}{2T}\right) \right) $ in Taylor series at small $m$.
This gives
\begin{eqnarray}
F &=&F_{0}+6\int_{0}^{\infty }\frac{{\cal B}_{1/2}^{H}(\pi t)e^{-{\cal F}%
_{0}(t/2T)}}{{\cal F}_{1}\left( \frac{t}{2T}\right) }dt+\frac{1}{2}%
m^{2}-m^{2}A_{1}^{H}+\frac{1}{20}m^{4}A_{3}^{H}+...,  \label{ap8} \\
A_{n}^{H} &=&\int_{0}^{\infty }{\cal B}_{1/2}^{H}(\pi t)e_{1}^{-{\cal F}%
_{0}(t/2T)}{\cal F}_{1}^{n}\left( \frac{t}{2T}\right) dt.  \nonumber
\end{eqnarray}
\end{widetext}
Paying attention that second term in (\ref{ap8}) does not depend on $m$ and
hence just renormalizes $F_{0}$, we easily obtain Eq. (\ref{q1}) from the
text.

\section{}
We are looking for the expression (\ref{eq5r}) for the case $S=5/2$. We
introduce notation ${\vec{t}}=J({\vec{r}}){\vec{\tau}}$ and assume $T=1$
(i.e. ${\vec{H}}$ means ${\vec{H}}/T$). In these notations Eq.(\ref{eq5r})
reads
\begin{equation}
W({\vec{H}})=1-\frac{{\rm Tr}\left\{ \exp \left[ i{\vec{S}}{\vec{t}}\right]
\exp \left[ -{\vec{S}}{\vec{H}}\right] \right\} }{{\rm Tr}\left\{ \exp \left[
-{\vec{S}}{\vec{H}}\right] \right\} }.  \label{ac1}
\end{equation}%
The denominator in Eq.(\ref{ac1}) can be immediately evaluated in a
reference frame rotating around quantization axis
\begin{equation}
{\rm Tr}\left\{ \exp \left[ -{\vec{S}}{\vec{H}}\right] \right\} =\frac{\sinh
3H}{\sinh H/2}.  \label{ac2}
\end{equation}%
We introduce the functions
\begin{eqnarray}
&&\phi _{n}=\frac{\cosh \left( nH\right) }{1+2\cosh \left( 2H\right) },
\label{ac3} \\
&&\eta _{n}=\frac{\cosh \left( nH\right) \left( \cosh H-1\right) }{\sinh 3H}=
\label{ac4} \\
&=&\frac{\cosh \left( nH\right) \sinh H/2}{\cosh H/2+\cosh 3H/2+\cosh 5H/2}
\nonumber
\end{eqnarray}%
and cosine of the angle between vectors ${\vec{H}}$ and ${\vec{\tau}}%
\parallel {\vec{t}}$,%
\begin{equation}
c=\cos \left( \widehat{{\vec{H}},{\vec{\tau}}}\right) =\frac{\left( {\vec{H}}%
{\vec{\tau}}\right) }{H\tau }.  \label{ac5}
\end{equation}%
In these notations after lengthy calculations we can get the
expression for the trace in numerator of Eq. (\ref{ac1}). With
respect to Eq. (\ref{ac2}), this expression assumes the form
\begin{widetext}
\begin{eqnarray}
W({\vec{H}}) &=&1-\frac{1}{4}\cos \frac{t}{2}\left\{ \left(
3-14c^{2}+15c^{4}\right) \phi _{0}-4\left( 1-6c^{2}+5c^{4}\right) \phi
_{1}+5\left( 1-c^{2}\right) ^{2}\phi _{2}\right\}-  \nonumber \\
&&-\frac{1}{8}\cos \frac{3t}{2}\left\{ \left( -1+38c^{2}-45c^{4}\right) \phi
_{0}+4\left( 1-12c^{2}+15c^{4}\right) \phi _{1}+5\left(
1+2c^{2}-3c^{4}\right) \phi _{2}\right\}-  \nonumber \\
&&-\frac{1}{8}\cos \frac{5t}{2}\left\{ \left( 3-10c^{2}+15c^{4}\right) \phi
_{0}+4\left( 1-5c^{4}\right) \phi _{1}+\left( 1+10c^{2}+5c^{4}\right) \phi
_{2}\right\}+  \nonumber \\
&&+\frac{ic}{4}\sin \frac{t}{2}\left\{ \left( 3-14c^{2}+15c^{4}\right) \eta
_{0}-4\left( 1-6c^{2}+5c^{4}\right) \eta _{1}+5\left( 1-c^{2}\right)
^{2}\eta _{2}\right\}-  \nonumber \\
&&-\frac{ic}{8}\sin \frac{3t}{2}\left\{ \left( 3-26c^{2}+15c^{4}\right) \eta
_{0}-4\left( 3-4c^{2}+5c^{4}\right) \eta _{1}-5\left( 3-2c^{2}-c^{4}\right)
\eta _{2}\right\}+  \nonumber \\
&&+\frac{ic}{8}\sin \frac{5t}{2}\left\{ \left( 15-10c^{2}+3c^{4}\right) \eta
_{0}+4\left( 5-c^{4}\right) \eta _{1}+\left( 5+10c^{2}+c^{4}\right) \eta
_{2}\right\} .  \label{ac6}
\end{eqnarray}
\end{widetext}


\end{document}